\documentclass[showpacs,twocolumn]{revtex4}
\usepackage{amssymb}
\usepackage{amsmath}
\usepackage{latexsym}
\usepackage{epsfig}

\begin{document}

\title{Anomalous reaction-transport processes: the dynamics beyond the Mass
Action Law}
\author{Daniel Campos$^{1}$, Sergei Fedotov$^{1}$ and Vicen\c{c} M\'{e}ndez$%
^{1,2}$ }
\affiliation{$^{1}$School of Mathematics, The University of Manchester, Manchester M60
1QD, UK.}
\affiliation{$^{2}$Grup de F\'{\i}sica Estad\'{\i}stica, Departament de F\'{\i}sica.
Facultat de Ci\`{e}ncies, Universitat Aut\`{o}noma de Barcelona, 08193
Bellaterra (Barcelona) Spain}

\begin{abstract}
In this paper we reconsider the Mass Action Law (MAL) for the
anomalous reversible reaction $A\rightleftarrows B$ with diffusion.
We provide a mesoscopic description of this reaction when the
transitions between two states $A$ and $B$ are governed by anomalous
(heavy-tailed) waiting-time distributions. We derive the set of
mesoscopic integro-differential equations for the mean densities of
reacting and diffusing particles in both states. We show that the
effective reaction rate memory kernels in these equations and the
uniform asymptotic states depend on transport characteristics such
as jumping rates. This is in contradiction with the classical
picture of MAL. We find that transport can even induce an extinction
of the particles such that the density of particles $A$ or $B$ tends
asymptotically to zero. We verify analytical results by Monte Carlo
simulations and show that the mesoscopic densities exhibit a
transient growth before decay.
\end{abstract}

\maketitle

The Mass Action Law (MAL) plays a very important role in a large number of
chemical, biological and physical systems \cite{kei,ma}. It states that the
rate of an elementary reaction is proportional to concentrations of
reactants. MAL also gives the expression for the equilibrium constant which
is a main characteristic of chemical equilibrium. To illustrate this,
consider the reversible reaction $A\rightleftarrows B$, with $\beta _{1}$
and $\beta _{2}$ denoting the forward and backward rate constants. According
to MAL, the balance equations for the mean concentrations $n_{1}$ and $n_{2}$
of diffusing particles $A$ and $B$ can be written as%
\begin{eqnarray}
\frac{\partial n_{1}}{\partial t} &=&D_{1}\nabla ^{2}n_{1}-\beta
_{1}n_{1}+\beta _{2}n_{2},  \notag \\
\frac{\partial n_{2}}{\partial t} &=&D_{2}\nabla ^{2}n_{2}+\beta
_{1}n_{1}-\beta _{2}n_{2},  \label{1}
\end{eqnarray}%
where $D_{1}$ and $D_{2}$ are the diffusion coefficients of the particles $A$
and $B$ respectively. It follows from (\ref{1}) that the uniform equilibrium
state $(n_{1}^{\infty },n_{2}^{\infty })$ obeys the equation
\begin{equation}
\frac{n_{1}^{\infty }}{n_{2}^{\infty }}=\frac{\beta _{2}}{\beta _{1}}=K_{eq},
\label{1b}
\end{equation}%
where $K_{eq}$ represents the equilibrium constant of the reaction process.
This constant depends on the thermodynamics properties of the system, but is
independent from the transport parameters $D_{1}$ and $D_{2}$. The purpose
of this paper is to reconsider these two fundamental equations (\ref{1}) and
(\ref{1b}) for anomalous reaction and transport.

Continuous time random walk (CTRW) models have been widely used in recent
years to gain insights into the anomalous transport \cite{kla}. The
extension of CTRW models to reaction-transport phenomena presents modeling
challenges, because of the difficulty of taking into account chemical
reactions within anomalous transport. Recently several authors have explored
the reaction-transport models in which the standard diffusion is replaced by
an anomalous (subdiffusive) transport \cite{Ka,
Se,fediom,soko,fedmen,yad,hen,hen2}. It has been shown that the evolution
equations for density of particles are drastically different from the
standard reaction-diffusion equations. For example, the transport and
reaction terms are not separable as it happens in the classical case (\ref{1}%
). Instead, one finds that the transport term becomes dependent on the
reaction constants $\beta _{1}$ or $\beta _{2}$ \cite{yad,fediom}. The
master equation for the mean density of one of reactants may include crossed
transport term \cite{soko,hen2}. This is a consequence of the non-Markovian
nature of subdiffusion.

In previous works, however, reaction was always introduced
phenomenologically following the principles of classical reaction kinetics.
The idea of this paper is to consider both the reversible reaction $%
A\rightleftarrows B$ and sub-diffusive transport from a probabilistic point
view. It is well known that the classical kinetics in (\ref{1}) corresponds
to Markovian transition of particles from one state to another. Our aim is
to take into account anomalous (non-Markovian) transitions of particles from
the state $A$ to $B$ and backward and find out how the transport process and
reactions are coupled. In what follows we will show that fundamental
constant $K_{eq}$ becomes dependent on transport characteristics which is in
contradiction to the classical picture of MAL. This is due to anomalous
nature of transitions $A\rightleftarrows B$ for which the waiting time
distribution exhibits the power law decay with the infinite mean waiting
time. Let us mention that the situations that are outside the scope of the
MAL have already been reported for diffusion-limited reactions with
long-range interactions in space \cite{voi}. The reversible reaction $%
A\rightleftarrows B$ can be interpreted as a switching between two states $A$
and $B$. This topic has attracted a great interest recently because the
switching process can be non-Markovian. Examples include two-state ion
channel gating \cite{Go}, stochastic resonance \cite{Go2}, quantum dots \cite%
{bro,bar}, etc. For anomalous switching process without transport the mean
residence time of the particles in each state is divergent \cite{shu,bro}.
As a result, the density of particles in one state tends to zero in the
limit $t\rightarrow \infty $ which means the extinction of one of the
states. Here we show that the transport process can drastically modify
extinction/survival dynamics for anomalous transitions. One of the
motivations for our study is the experimental data for a malignant brain
cancer that exhibits migration-proliferation dichotomy \cite{Giese2}. The
motility (transport) of cells and phenotype transitions $A\rightleftarrows B$
(proliferation$\rightleftharpoons $migration) can be anomalous
simultaneously \cite{fediom}. Another possible application of our model is
the isomerization reaction for which macromolecule in two interconvertible
states migrate with different electrophoretic mobilities \cite{Cann}.

In this Letter we consider the following stochastic model for the transport
and reversible reaction $A\rightleftarrows B$. The particles of type $A$ and
$B$ randomly move along one-dimensional space and switch between the states $%
A$ and $B$. This random walk with switching can be described by four
sequences of mutually independent random variables. Two sequences $\left\{
Y_{1,}Y_{2,...}\right\} $ and $\left\{ Z_{1,}Z_{2,...}\right\} $ describe
the waiting times between jumps for particles in the state $A$ and $B$
correspondingly. We assume that these random variables are identically
distributed with probability density function (pdf) $\varphi _{1}(t)$ for
particles $A$ and \ pdf $\varphi _{2}(t)$ for particles $B$. Two other
sequences $\left\{ U_{1,}U_{2,...}\right\} $ and $\left\{
W_{1,}W_{2,...}\right\} $ describe the waiting times for random transitions:
$A\rightarrow B$ and $B\rightarrow A$ respectively. The random variables $%
U_{1,}U_{2,...}$ and $W_{1,}W_{2,...}$are identically distributed with pdf's
$\psi _{12}(t)$ and $\psi _{21}(t)$. If we place the particle at the
position $x$ at time $0$ in state $B,$ and if the random time $Z_{1}$ for a
jump is less than random time $W_{1}$ for reaction $B\rightarrow A$, then
the random jump happens at time $Z_{1}.$ However if $W_{1}<Z_{1}$, then the
transition $B\rightarrow A$ occurs at time $W_{1}.$ In other words an event
(reaction or jump) happens at time $\min \left( W_{1},Z_{1}\right) .$ For
example, there is a jump at the position $X_{1}$ at time $Z_{1}<W_{1},$ then
a second jump of length $X_{2}$ after a further time $Z_{2}<W_{2},$ a
switching to state $A$ after time $W_{3}<Z_{3},$ a transition back to $B$
after time $U_{4}<Y_{4},$ and so on (see Fig.1). So jumps and transitions $%
A\rightleftarrows B$ are not independent random events as in \cite{soko,hen2}%
.

We express the density of particles $n_{i}(x,t)$ at position $x$ at time $t$
in terms of the initial distribution of particles $n_{i}^{0}(x)$ and density
of particles $j_{i}(x,t-\tau )$ that arrive at the same position $x$ at
previous time $t-\tau .$ The balance equations for the mean densities $n_{1}$
and $n_{2}$ of particles $A$ and $B$ are
\begin{eqnarray}
n_{1}(x,t) &=&n_{1}^{0}(x)\Phi _{1}(t)\Psi _{12}(t)+  \notag \\
&&\int_{0}^{t}j_{1}(x,t-\tau )\Phi _{1}(\tau )\Psi _{12}(\tau )d\tau ,
\notag \\
n_{2}(x,t) &=&n_{2}^{0}(x)\Phi _{2}(t)\Psi _{21}(t)+  \notag \\
&&\int_{0}^{t}j_{2}(x,t-\tau )\Phi _{2}(\tau )\Psi _{21}(\tau )d\tau ,
\label{balance}
\end{eqnarray}%
where $\Phi _{i}(t)=\int_{t}^{\infty }\varphi _{i}(\tau )d\tau $ and $\Psi
_{ij}(t)=\int_{t}^{\infty }\psi _{ij}(\tau )d\tau $ are the corresponding
survival probabilities for $\varphi _{i}(t)$ and $\psi _{ij}(t)$. For
example, $\Phi _{1}(t)$ is the probability that a particle in the state $A$
does not jump until time $t$, and $\Psi _{12}(t)$ is the probability that a
particle in the state $A$ does not switch to $B$ until time $t$. The first
term in the RHS of the Eq. (\ref{balance}) represents the contribution from
the initial density of particles that have neither jumped nor switched until
time $t$. The density $j_{i}(x,t)$ describes how the particles arrive at
point $x$ at time $t$ as a result of the transport and switching processes.
Equations for the density $j_{1}(x,t)$ of particles $A$ and the density $%
j_{2}(x,t)$ of particles $B$ can be written as
\begin{align}
j_{1}(x,t)& =n_{2}^{0}(x)\Phi _{2}(t)\psi _{21}(t)+  \notag \\
& \varphi _{1}(t)\Psi _{12}(t)\int_{-\infty }^{\infty }n_{1}^{0}(x-z)\rho
_{1}(z)dz+  \notag \\
& \int_{0}^{t}j_{2}(x,t-\tau )\Phi _{2}(\tau )\psi _{21}(\tau )d\tau +
\notag \\
& \int_{0}^{t}\int_{-\infty }^{\infty }j_{1}(x-z,t-\tau )\varphi _{1}(\tau
)\Psi _{12}(\tau )\rho _{1}(z)dzd\tau  \notag \\
j_{2}(x,t)& =n_{1}^{0}(x)\Phi _{1}(t)\psi _{12}(t)+  \notag \\
& \varphi _{2}(t)\Psi _{21}(t)\int_{-\infty }^{\infty }n_{2}^{0}(x-z)\rho
_{2}(z)dz+  \notag \\
& \int_{0}^{t}j_{1}(x,t-\tau )\Phi _{1}(\tau )\psi _{12}(\tau )d\tau +
\notag \\
& \int_{0}^{t}\int_{-\infty }^{\infty }j_{2}(x-z,t-\tau )\varphi _{2}(\tau
)\Psi _{21}(\tau )\rho _{2}(z)dzd\tau .  \label{3}
\end{align}%
The first equation is the conservation law for particles of type $A$ at time
$t$ at position $x.$ The first term in the RHS of the equation accounts for
the initial distribution $n_{2}^{0}(x)$ of particles in state $B$ that
switch to $A$ at time $t$, provided they do not jump up to time $t$. The
second term is the contribution from the initial density $n_{1}^{0}(x)$ of
particles in state $A$ that jump to $x$ from $x-z$ at time $t$ having not
switched until $t$. The random jump length $z$ is determined by the
dispersal kernel $\rho _{1}(z)$. The third term represents the contribution
from those particles that switch from state $B$ to state $A$ after a waiting
time $\tau $, under the condition that they do not jump during that time.
Finally, the fourth term corresponds to the contribution of particles in
state $A$ that arrive at $x-z$ at $t-\tau $ do not switch to $B$ during time
$\tau $.

\begin{figure}[tbp]
\begin{center}
\epsfxsize=5cm \leavevmode \epsffile{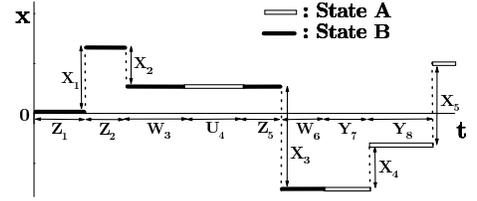}
\end{center}
\caption{A single realization of the random walk with switching $%
A\rightleftarrows B$. Initially the particle is assumed to be in the state $%
B $ and at position $x=0$. }
\end{figure}

The set of linear equations (\ref{balance}),(\ref{3}) can be solved by using
the Laplace-Fourier transforms: $\left( x,t\right) \rightarrow \left(
q,s\right) $. One can obtain two equations: $sn_{1}(q,s)-n_{1}^{0}(q)=k_{1}%
\left( s\right) (\rho _{1}\left( q\right) -1)n_{1}(q,s)-a_{1}\left( s\right)
n_{1}(q,s)+a_{2}\left( s\right) n_{2}(q,s)$ and $%
sn_{2}(q,s)-n_{2}^{0}(q)=k_{2}\left( s\right) (\rho _{2}\left( q\right)
-1)n_{2}(q,s)+a_{1}\left( s\right) n_{1}(q,s)-a_{2}\left( s\right)
n_{1}(q,s).$ Here we introduce
\begin{equation}
k_{i}\left( s\right) \equiv \frac{\left[ \varphi _{i}\Psi _{ij}\right] _{s}}{%
\left[ \Phi _{i}\Psi _{ij}\right] _{s}}\qquad a_{i}\left( s\right) \equiv
\frac{\left[ \Phi _{i}\psi _{ij}\right] _{s}}{\left[ \Phi _{i}\Psi _{ij}%
\right] _{s}}.  \label{6}
\end{equation}%
with the notation $\left[ f\right] _{s}\equiv f\left( s\right) $. Taking the
inverse Laplace and Fourier transforms we obtain the following master
equations
\begin{align}
\frac{\partial n_{1}}{\partial t}& =\int_{0}^{t}\int_{-\infty }^{\infty
}n_{1}(t-\tau ,x-z)k_{1}(\tau )(\rho _{1}(z)-1)dzd\tau +  \notag \\
& -\int_{0}^{t}a_{1}(\tau )n_{1}(t-\tau ,x)d\tau +\int_{0}^{t}a_{2}(\tau
)n_{2}(t-\tau ,x)d\tau  \notag \\
\frac{\partial n_{2}}{\partial t}& =\int_{0}^{t}\int_{-\infty }^{\infty
}n_{2}(t-\tau ,x-z)k_{2}(\tau )(\rho _{2}(z)-1)dzd\tau +  \notag \\
& +\int_{0}^{t}a_{1}(\tau )n_{1}(t-\tau ,x)d\tau -\int_{0}^{t}a_{2}(\tau
)n_{2}(t-\tau ,x)d\tau ,  \label{6b}
\end{align}%
where $k_{i}(t)$ and $a_{i}(t)$ are the inverse Laplace transforms of $%
k_{i}\left( s\right) $, $a_{i}\left( s\right) $ defined in
(\ref{6}). The most interesting feature of the system (\ref{6b}) is
that effective reaction rate memory kernels $a_{1}(t)$ and
$a_{2}(t)$ depend on the transport through the survival
probabilities $\Phi _{1}(t)$ and $\Phi _{2}(t),$ while the transport
memory kernels $k_{1}(t)$ and $k_{2}(t)$ depend on statistical
characteristics of reactions such as $\psi _{ij}.$ If the random
waiting times for switching and jumping are exponentially
distributed: $\varphi _{i}=\lambda _{i}e^{-\lambda _{i}t}$, $\psi
_{ij}=\beta _{i}e^{-\beta _{i}t}, $ then these dependencies cease to
exist. As a result the transport and reaction terms are separable as
in the classical case (\ref{1}). For example, if we use the
diffusive approximation for transport $\rho _{i}\left( q\right) \sim
1-\sigma _{i}^{2}q^{2}$, then the system (\ref{6b})
can be written as classical reaction-diffusion equations (\ref{1}), with $%
D_{i}\equiv \lambda _{i}\sigma _{i}^{2}$. Similarly, for a Markovian
switching process with subdiffusive transport, we could recover from (\ref%
{6b}) the model for cancer spreading studied in \cite{fediom}. If the
waiting time pdf $\psi _{12}$ has a gamma distribution as $\psi _{12}=\beta
_{1}^{2}te^{-\beta _{1}t}$ and $\varphi _{1}=\lambda _{1}e^{-\lambda _{1}t},$
then $a_{1}\left( s\right) =\beta _{1}^{2}\left( 2\beta _{1}+\lambda
_{1}+s\right) ^{-1}$. So the reaction rate memory kernels are
\begin{equation}
a_{i}(\tau )=\beta _{i}^{2}e^{-\left( 2\beta _{i}+\lambda _{i}\right) \tau
}\qquad i=1,2.
\end{equation}%
This formula shows that the effective reaction rate kernels depend on the
rate of jumps $\lambda _{i}$. Now, let us find the uniform stationary states
corresponding to (\ref{6b}) under the condition $n_{1}^{\infty
}+n_{2}^{\infty }=1$. From the limit $q\rightarrow 0$ one finds $\rho
_{i}\left( q\right) =1,$ so we obtain the asymptotic state as $s\rightarrow
0 $
\begin{equation}
(n_{1}^{\infty },n_{2}^{\infty })=\lim_{s\rightarrow 0}\left( \frac{%
a_{2}\left( s\right) }{a_{1}\left( s\right) +a_{2}\left( s\right) },\frac{%
a_{1}\left( s\right) }{a_{1}\left( s\right) +a_{2}\left( s\right) }\right) .
\label{7}
\end{equation}%
The main feature of this asymptotic state is that in general it depends on
the characteristics of the transport process which is in contradiction with
the Mass Action Law (\ref{1b}). This follows from the fact that the survival
function $\Phi _{i}$ appears in the definition of $a_{i}$. This happens for
any situation except when the switching process is Markovian for which $%
\lim_{s\rightarrow 0}a_{i}=\beta _{i}$.

Assume now that the reaction process is governed by a power-law decaying
distribution of waiting times. We use the standard approximation $\left[
\psi _{ij}\right] _{s}\sim 1-\left( \beta _{i}^{-1}s\right) ^{\gamma _{ij}}$
with $\gamma _{ij}<1$ as $s\rightarrow 0$. On the contrary, for the
transport we consider the Markovian case: $\varphi _{i}(t)=\lambda
_{i}e^{-\lambda _{i}t}$. Then, we find that the state (\ref{7}) turns into
\begin{equation}
\left( \frac{\lambda _{2}^{1-\gamma _{21}}\beta _{2}^{\gamma _{21}}}{\lambda
_{1}^{1-\gamma _{12}}\beta _{1}^{\gamma _{12}}+\lambda _{2}^{1-\gamma
_{21}}\beta _{2}^{\gamma _{21}}},\frac{\lambda _{1}^{1-\gamma _{12}}\beta
_{1}^{\gamma _{12}}}{\lambda _{1}^{1-\gamma _{12}}\beta _{1}^{\gamma
_{12}}+\lambda _{2}^{1-\gamma _{21}}\beta _{2}^{\gamma _{21}}}\right) ,
\label{8}
\end{equation}%
where the explicit dependencies of asymptotic states on the transport
parameters $\lambda _{1}$ and $\lambda _{2}$ are evident. In fact, the ratio
of the two uniform densities in the limit $t\rightarrow \infty $ is
\begin{equation}
\frac{n_{1}^{\infty }}{n_{2}^{\infty }}=\frac{\lambda _{2}^{1-\gamma
_{21}}\beta _{2}^{\gamma _{21}}}{\lambda _{1}^{1-\gamma _{12}}\beta
_{1}^{\gamma _{12}}}=K,  \label{8b}
\end{equation}%
where $K$ becomes dependent on the transport parameters $\lambda _{1}$ and $%
\lambda _{2}$ (see (\ref{1b}) for comparison). Note, however, this
new constant $K$ cannot be considered an equilibrium constant since
in the non-Markovian case considered here a thermodynamic
equilibrium state cannot be defined. Let us assume $\beta _{1}<\beta
_{2}$; then the MAL (\ref{1b}) predicts $n_{1}^{\infty
}>n_{2}^{\infty }$. On the contrary, it is clear from (\ref{8b})
that one could choose the rates of jumping $\lambda _{1}$ and
$\lambda _{2}$ so that the inequality can be inverted to
$n_{1}^{\infty }<n_{2}^{\infty }$. We obtain even more dramatic
results if we take the limit $\lambda _{1}\rightarrow 0$ or $\lambda
_{2}\rightarrow 0$. If, for example, we consider the limit $\lambda
_{1}\rightarrow 0$ $\left( \lambda _{2}\neq 0\right) $, then one can
observe the extinction of particles in state $B$ and survival of
particles in state $A$, that is, $n_{1}\rightarrow 1$ and
$n_{2}\rightarrow 0$ as $t\rightarrow \infty $ \ (see (\ref{8})). So
we find from our model that transport process can induce a
survival/extinction of one of the two densities for anomalous
reactions. To validate this phenomenon we have performed the direct
Monte Carlo simulations of two-states random walks. The results are
illustrated in Fig. 2 where one can see that if $\lambda _{1}\neq 0$
and $\lambda _{2}=0,$ then we might observe the temporal growth of
$n_{1}$ before the final decay to zero (solid line). However, if we
put $\lambda _{1}=\lambda _{2}=0$ (dashed line),  then the limit for
the density of particles in state $A$ is completely different, that
is,  $n_{1}\rightarrow 1$ as $t\rightarrow \infty .$

\begin{figure}[tbp]
\begin{center}
\epsfxsize=5.0cm \leavevmode \epsffile{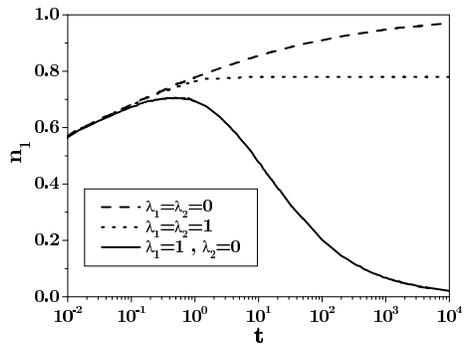}
\end{center}
\caption{Time evolution of the density $n_{1}$ obtained from Monte Carlo
simulations for different values of the parameters $\protect\lambda _{1}$
and $\protect\lambda _{2}$; $\protect\sigma _{1}^{2}=\protect\sigma %
_{2}^{2}=1$, $\protect\beta _{1}=\protect\beta _{2}=1 $, $\protect\gamma %
_{12}=0.25$, $\protect\gamma _{21}=0.5$. }
\end{figure}

The result $n_{i}\rightarrow 1$ as $\lambda _{i}\rightarrow 0$ imply that if
the particles do not move in one of the states, they survive. This idea of
\textit{'staying quiet helps you to survive'} can be understood from the
interplay between the waiting times for reactions and jumps. According to
our derivation, the reaction process, say the transition from state $A$ to
state $B$, is actually governed by the density $\Phi _{1}\psi _{12}$ (the
particles react only if they have not jumped before, as can be seen from (%
\ref{3})). We can refer to this function as the \textit{effective} waiting
time pdf. For an anomalous switching process with Markovian transport the
asymptotic behavior of the \textit{effective} pdf reads $\Phi _{1}\psi
_{12}\sim t^{-1-\gamma _{12}}e^{-\lambda _{1}t}$. Then, the mean waiting
time is finite, and for this reason the system reaches a stationary state,
given by (\ref{8}). However, in the limit $\lambda _{1}\rightarrow 0$ the
\textit{effective} mean waiting time diverges, which makes the reaction $%
A\rightarrow B$ much slower than the backward reaction $B\rightarrow A$, so
the particles tend to get trapped in the state $A$. For this reason we
obtain $n_{1}\rightarrow 1$ and $n_{2}\rightarrow 0$ in the long-time limit.

\begin{figure}[tbp]
\begin{center}
\epsfxsize=5.0cm \leavevmode \epsffile{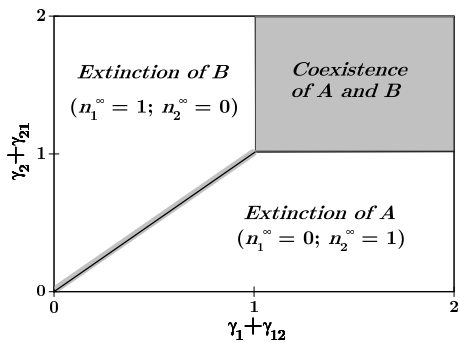}
\end{center}
\caption{ Phase diagram of the extinction/survival regions for anomalous
switching and transport in terms of the exponents $\protect\gamma _{i}$ and $%
\protect\gamma _{ij}$. }
\end{figure}

For anomalous transport and anomalous reaction (switching), we have $\psi
_{ij}(s)\sim 1-\left( \beta _{i}^{-1}s\right) ^{\gamma _{ij}}$ and $\varphi
_{i}(s)\sim 1-\left( \lambda _{i}^{-1}s\right) ^{\gamma _{i}}$, with $\gamma
_{i}<1$ and $\gamma _{ij}<1$. It is helpful again to use the idea of an
\textit{effective} waiting time distribution for reaction: $\Phi _{i}\psi
_{ij}\sim t^{-1-\gamma _{i}-\gamma _{ij}}$. So that, the \textit{effective}
mean waiting time can be finite, provided that the condition $\gamma
_{i}+\gamma _{ij}>1$ is satisfied. Fig. 3 shows the "phase diagram" for
asymptotic states $n_{1}^{\infty }$ and $n_{2}^{\infty }$ depending of the
values of $\gamma _{1}+\gamma _{12}$ and $\gamma _{2}+\gamma _{21}$. If both
of them are larger than one, then the transitions $1\rightarrow 2$ and $%
2\rightarrow 1$ are governed by finite \textit{mean} waiting times, so a
coexistence of two states is possible. In other regions, the divergences of
the \textit{mean} waiting times make the particles get trapped in the state
where the switching process is slower. So that, one of the states become
extinct in the asymptotic regime. These results can be explained by a
coupled renewal property assumed in our model. If the "internal" waiting
time of the particles starts from zero after each event (reaction or jump),
then we have a competition between both processes to be the first to occur,
and so coupled effects emerge. This coupled renewal property is opposite to
additive renewal property when the random walk in space is completely
independent of the reaction process (see, for example, \cite{soko}).

To sum up, we have presented a non-Markov model for the reversible reaction $%
A\rightleftarrows B$ and studied the interplay between anomalous transport
and anomalous reaction process implemented in a probabilistic way. So we
have been able to explore those situations that are beyond Mass Action Law.
We have derived mesoscopic integro-differential equations for the mean
densities of particles in states $A$ and $B$ when the transitions between
two states $A$ and $B$ and jumps in space are governed by heavy-tailed
waiting-time distributions. It has been shown that the anomalous properties
of the reversible reaction yield the appearance of unusual properties such
as dependence of asymptotic states on transport and the transient growth of
densities before decay. We have found that the transport process can modify
completely the uniform stationary regimes. In particular, it can induce the
survival/extinction of one of the states. These results have been validated
by direct Monte Carlo simulations of two-state random walk.

\textbf{Acknowledgements.} This research has been partially supported by the
Generalitat de Catalunya by the grant 2006-BP-A-10060 (DC), and by Grants
Nos. FIS 2006-12296-C02-01, SGR 2005-00087 (VM) and EPSRC EP/D03115X/1 (SF
and VM).

\end{document}